\documentstyle[12pt]{article}
\newcommand{\bce}{\begin{center}}
\newcommand{\ece}{\end{center}}
\newcommand{\beq}{\begin{equation}}
\newcommand{\eeq}{\end{equation}}
\newcommand{\bea}{\vspace{0.25cm}\begin{eqnarray}}
\newcommand{\eea}{\end{eqnarray}}

\newcommand{\ba}{\begin{array}}
\newcommand{\ea}{\end{array}}

\newcommand{\r}{\mbox{{\boldmath $\rho$}}}

\newcommand{\doublespace}{
    \renewcommand{\baselinestretch}{1.6}\large\normalsize}

\def\lsim{\mathrel{\rlap{\lower4pt\hbox{\hskip1pt$\sim$}}
    \raise1pt\hbox{$<$}}}         %less than or approx. symbol
\def\gsim{\mathrel{\rlap{\lower4pt\hbox{\hskip1pt$\sim$}}
    \raise1pt\hbox{$>$}}}         %greater than or approx. symbol

\def\beq{\begin{equation}}
\def\endeq{\end{equation}}
\def\bea{\begin{eqnarray}}
\def\arr{\begin{eqnarray}}
\def\eea{\end{eqnarray}}

\def\q2{$Q^{2}$}
\def\s2{2$S$}

\begin{document}
\doublespace 
\begin{titlepage}
\vspace*{-2cm}
\begin{flushright}
{\large\bf
KFA-IKP(TH)-1996-17\\
\,\,30 October 1996\,\,\,\,\,\,\,\,}
\end{flushright}
 
\bigskip
 
\begin{center}
  
  {\huge\bf

    Landau--Pomeranchuk--Migdal effect
for  finite-size targets
}
 
  \vspace{1.5 cm}
 
  {\Large    B. G. Zakharov }
  \bigskip
  \bigskip

{\it  Institut  f\"ur Kernphysik,
        Forschungszentrum J\"ulich,\\
        D-52425 J\"ulich, Germany\medskip\\
 L. D. Landau Institute for Theoretical Physics,
        GSP-1, 117940,\\ ul. Kosygina 2, 117334 Moscow, Russia
        \medskip\\}

\vspace{.5cm}

  {\bf\large Abstract}
\end{center}
A rigorous evaluation of the Landau-Pomeranchuk-Migdal 
effect for finite-size targets
is performed  within the 
path integral approach previously developed in ref. \cite{BGZLPM}.
The bremsstrahlung rate in QED is expressed through a solution
of a two-dimensional Schr\"odinger equation with an imaginary potential.
The boundary condition for this solution is formulated in terms
of a product of the light-cone
electron--photon wave function and the dipole cross section for scattering 
of $e^{+}e^{-}$ pair off an atom.
Numerical calculations are performed for homogeneous and structured targets.
Our predictions for the  homogeneous target 
agree well with the photon spectrum measured recently at SLAC with
25 GeV electrons.
The spectra obtained for the structured two segment targets exhibit the
interference minima and maxima.

\vspace*{3cm}
 
\noindent

%\vfill
%\newpage
\end{titlepage}
 
%*****************************************************

\newpage
%\pacs{PACS numbers: 13.60.Le, 12.38.Qk, 25.20.Lj}

%\narrowtext
%\twocolumn
%\doublespace
{\bf 1.}
Forty years ago Migdal \cite{Migdal} developed a 
quantum theory of suppression of the bremsstrahlung
rate in a dense medium predicted by Landau and
Pomeranchuk \cite{LP}. The first accurate measurement 
of the Landau-Pomeranchuk-Migdal
(LPM) effect was recently performed at SLAC \cite{SLAC}.
Although the experiment \cite{SLAC} corroborated the 
suppression of the radiation rate predicted in refs. \cite{LP,Migdal},
for sufficiently thin targets the measured spectra
disagree with prediction of Migdal's theory. This disagreement
is presumably connected with neglecting the edge effects in ref.
\cite{Migdal}, where the case of an infinite medium was considered.
Besides in ref. \cite{Migdal} 
the radiation rate for the infinite
medium was calculated under certain
approximations. Namely, Migdal used the Fokker-Planck approximation for
evaluation of the electron density matrix. Furthermore,
the inelastic processes 
with excitations of atoms were neglected. 
For these reasons Migdal's approach does not reproduce
the Bethe-Heitler spectrum in a limit of low target density 
when the LPM suppression vanishes.

In the present paper we evaluate the LPM effect for 
finite-size homogeneous and structured targets 
within the approach developed in ref. \cite{BGZLPM}.
This approach is based on the path integral treatment of multiple
scattering of ref. \cite{BGZA2}, and a new formulation
of the light-cone perturbation theory in terms of transverse
Green's functions. 
Contrary to ref. \cite{Migdal} the approach \cite{BGZLPM} 
treats  the evolution of the electron density 
matrix rigorously, and allows inclusion of inelastic processes.
Within the normalization factor $\sim 0.93$ our results agree well with 
the experimental data \cite{SLAC} obtained for 25 GeV
electron beam interacting with a homogeneous gold target.
For structured targets we predict interference minima
and maxima in the photon spectra.
%***********************************************************

{\bf 2.}
In ref. \cite{BGZLPM} we reduced evaluation of the
bremsstrahlung rate to solving a two-dimensional Schr\"odinger equation
in the impact parameter space, for which the longitudinal coordinate 
$z$ plays the role of time and the Hamiltonian reads 
\beq
H=\frac{{\mbox{\bf{p}}}^2}{2\mu(x)}+v(\r,z)\,,
\label{eq:1}
\eeq
Here 
$
v(\r,z)=-i{n(z)\sigma(\rho x)}/{2}\,,
$
$
\mu(x)=E_{e}x(1-x)\,,
$
where $\sigma(\rho)$ is the dipole cross section for interaction
of $e^{+}e^{-}$ pair of size $\rho$ with an atom, $n(z)$ is the
target density, which is assumed to be independent of the transverse
coordinate $\r$, $E_{e}$ is the incident electron energy, $x=k/P_{e}$ 
is the Feynman
variable for the radiated photon.
The probability of photon radiation obtained in ref. \cite{BGZLPM} 
is given by 
\beq
\frac{d P}{d x}=2\mbox{Re}\!
\int\limits_{-\infty}^{\infty}\! d \xi_{1}\!
\int\limits_{\xi_{1}}^{\infty}d \xi_{2}
\exp\left[-\frac{i(\xi_{2}-\xi_{1})}{l_{f}}\right]
g(\xi_{1},\xi_{2},x)\left[K(0,\xi_{2}|0,\xi_{1})
-K_{v}(0,\xi_{2}|0,\xi_{1})\right]\,,
\label{eq:2}
\eeq
where $K$
is the Green's function for the Hamiltonian (\ref{eq:1}), $K_{v}$
is the vacuum Green's function, 
$
l_{f}=2E_{e}(1-x)/m_{e}^{2}x 
$ is the so called photon formation length.
The vertex operator $g(\xi_{1},\xi_{2},x)$, accumulating spin effects
in the transitions $e\rightarrow e'\gamma \rightarrow e$, is given by
\beq
g(\xi_{1},\xi_{2},x)=\Lambda_{nf}(x)
\frac{{\mbox{\bf p}}(\xi_{2})\cdot{\mbox{\bf p}}(\xi_{1})}{\mu^{2}(x)}
+\Lambda_{sf}(x)\,,
\label{eq:3}
\eeq
where
$\Lambda_{nf}(x)={\alpha[4-4x+2x^{2}]}/{4x}$,
$\Lambda_{sf}(x)=
\alpha m_{e}^{2}x[2E_{e}^{2}(1-x)^{2}]^{-1}$\,.
The two terms in (\ref{eq:3}) correspond to the transitions 
conserving (nf) and changing (sf) the electron helicity.

For numerical evaluation of the radiation rate it is convenient to
 rewrite Eq. (\ref{eq:2})
in another form. Expanding $K$ in a series
in the potential $v$  
$$
K(\r_{2},z_{2}|\r_{1},z_{1})=K_{v}(\r_{2},z_{2}|\r_{1},z_{1})
+\int\limits_{z_{1}}^{z_{2}}dz\int d\r 
K_{v}(\r_{2},z_{2}|\r,z)v(\r,z)
K_{v}(\r,z|\r_{1},z_{1})\,+\cdots,
$$
after a simple algebra one can represent (\ref{eq:2}) in the form
\beq
\frac{d P}{d x}=
\frac{d P_{BH}}{d x}+\frac{d P_{abs}}{d x}\,,
\label{eq:4}
\eeq
where
\bea
\frac{d P_{BH}}{d x}=-T\cdot\mbox{Re}
\int d\r
\int\limits_{-\infty}^{0} d\xi_{1}\int\limits_{0}^{\infty} d\xi_{2}
g(\xi_{1},\xi_{2},x)
K_{v}(0,\xi_{2}|\r,0)\nonumber\\
\times\sigma(\rho x)K_{v}(\r,0|0,\xi_{1})
\exp\left[-\frac{i(\xi_{2}-\xi_{1})}{l_{f}}\right]\,,
\label{eq:5}
\eea
\bea
\frac{d P_{abs}}{d x}=\frac{1}{2}\mbox{Re}
\int\limits_{0}^{L} dz_{1}n(z_{1})
\int\limits_{z_{1}}^{L} dz_{2}n(z_{2})
\int d\r_{1}d\r_{2}
\int\limits_{-\infty}^{0} d\xi_{1}\int\limits_{0}^{\infty} d\xi_{2}
g(\xi_{1},\xi_{2},x)
K_{v}(0,\xi_{2}|\r_{2},z_{2})\nonumber\\
\times\sigma(\rho_{2} x)K(\r_{2},z_{2}|\r_{1},z_{1})
\sigma(\rho_{1}x)
K_{v}(\r_{1},z_{1}|0,\xi_{1})
\exp\left[-\frac{i(\xi_{2}-\xi_{1}+z_{2}-z_{1})}{l_{f}}\right]\,.
\label{eq:6}
\eea
Here 
$
T=\int_{0}^{L} dz n(z)
$
is the optical thickness of the target (we assume that $n(z)=0$ at $z<0$
and $z>L$).
The integrals over $\xi_{1,2}$ in (\ref{eq:5}), (\ref{eq:6})
of the products of the vacuum Green's functions and exponential 
phase factors can be expressed through the light-cone 
wave function 
$
\Psi(x,\r,\lambda_{e},\lambda_{e'},\lambda_{\gamma})
$
for the transition $e\rightarrow e'\gamma$. At $\lambda_{e'}=\lambda_{e}$ 
it is 
\bea
\Psi(x,\r,\lambda_{e},\lambda_{e'},\lambda_{\gamma})=
\frac{-i}{2\mu(x)}\sqrt{\frac{\alpha}{2x}}
\left[\lambda_{\gamma}(2-x)+2\lambda_{e}x\right]
\left(\frac{\partial}{\partial \rho_{x}}-
i\lambda_{\gamma}\frac{\partial}{\partial \rho_{y}}\right)
\int\limits_{-\infty}^{0}d\xi K_{v}(\r,0|0,\xi)
\nonumber \\ \times
\exp\left(\frac{i\xi}{l_{f}}\right)=
\frac{1}{2\pi}\sqrt{\frac{\alpha x}{2}}
\left[\lambda_{\gamma}(2-x)+2\lambda_{e}x\right]
\exp(-i\lambda_{\gamma}\varphi)m_{e}K_{1}(\rho m_{e}x)\,,
\label{eq:7}
\eea
for $\lambda_{e'}=-\lambda_{e}$ the only nonzero
component is the one with $\lambda_{\gamma}=2\lambda_{e}$ 
\bea
\Psi(x,\r,\lambda_{e},-\lambda_{e},2\lambda_{e})=
\frac{\sqrt{2\alpha x^{3}}}{2\mu(x)}
\int\limits_{-\infty}^{0} d\xi K_{v}(\r,0|0,\xi)
\exp\left(\frac{i \xi}{l_{f}}\right)
=\frac{-i}{2\pi}\sqrt{2\alpha x^{3}}m_{e}K_{0}(\rho m_{e}x)\,.
\label{eq:8}
\eea 
Here $\alpha=1/137$, $K_{0}$ and $K_{1}$ are the Bessel functions.

Making use of Eqs.~(\ref{eq:7}),~(\ref{eq:8}) one can rewrite 
(\ref{eq:5}),~(\ref{eq:6}) in the form
\beq
\frac{d P_{BH}}{d x}=\frac{T}{2}\sum\limits_{\{\lambda_{i}\}}
\int d\r\,
|\Psi(x,\r,\{\lambda_{i}\})|^{2}
\sigma(\rho x)\,\,,
\label{eq:9}
\eeq
\bea
\frac{d P_{abs}}{d x}=-\frac{1}{4}\mbox{Re}
\sum\limits_{\{\lambda_{i}\}}
\int\limits_{0}^{L} dz_{1}n(z_{1})
\int\limits_{z_{1}}^{L} dz_{2}n(z_{2})
\int d\r\,
\Psi^{*}(x,\r,\{\lambda_{i}\})\nonumber\\
\times\sigma(\rho x)
\Phi(x,\r,\{\lambda_{i}\},z_{1},z_{2})
\exp\left[-\frac{i(z_{2}-z_{1})}{l_{f}}\right]\,,
\label{eq:10}
\eea  
where
\beq
\Phi(x,\r,\{\lambda_{i}\},z_{1},z_{2})=
\int d\r' K(\r,z_{2}|\r',z_{1})
\Psi(x,\r',\{\lambda_{i}\})\,\sigma(\rho' x)
\label{eq:11}
\eeq
is the solution of the Schr\"odinger equation with the boundary
condition   
$
\Phi(x,\r,\{\lambda_{i}\},z_{1},z_{1})=
\Psi(x,\r,\{\lambda_{i}\})\sigma(\rho x)\,.
$

In ref. \cite{NPZ} it was shown that the $p_{\perp}$-integrated
cross section   
for a process $a\rightarrow b c$
can be written as
\beq
\frac{d\sigma ({a\rightarrow cb})}{dx}=
\int d\r \,W_{a}^{bc}(x,\r)
\sigma_{\bar{a}bc}(\rho)\,,
\label{eq:12}
\eeq
where $W_{a}^{bc}$ is the light-cone probability distribution for 
transition $a\rightarrow bc$, $\sigma_{\bar{a}bc}$ is the total
cross section of interaction with the target of $\bar{a}bc$ system.
For the transition $e\rightarrow e'\gamma$ the corresponding 
three-body cross section equals $\sigma(\rho x)$. Thus, we see
that the first term in (\ref{eq:4}) equals the Bethe-Heitler cross
section times the target optical thickness, {\it i.e.} it
corresponds to the impulse approximation, 
while the second term describes the LPM suppression.
It is worth noting that at $l_{f}\gg L$ the whole radiation
rate can be also represented in the form analogous to Eq. (\ref{eq:12}).
Indeed, in this limit the transverse variable $\r$ is approximately
frozen, and the Green's function can be written
in the eikonal form  
\beq
K(\r_{2},z_{2}|\r_{1},z_{1})\approx 
\delta(\r_{2}-\r_{1})\exp
\left[-\frac{T\sigma(\rho_{1}x)}{2}\right]\,.
\label{eq:13}
\eeq
Making use of (\ref{eq:13}) we obtain in the frozen-size approximation 
\beq
\frac{dP_{fr}}{dx}=
2\int d\r \,W_{e}^{e\gamma}(x,\r)
\left\{1-\exp\left[-\frac{T\sigma(\rho x)}{2}\right]\right\}\,.
\label{eq:14}
\eeq
Eq. (\ref{eq:14}) is analogous to the formula for the cross section of heavy
quark production in hadron nucleus collision obtained 
in ref. \cite{NPZ}. In QED the LPM effect at $l_{f}\gg L$ was
previously discussed within soft photon approximation in ref. \cite{HOHLY}. 

%*****************************************************************
{\bf 3.}
The dominating values of $\rho$ in (\ref{eq:9}) are $\sim 1/m_{e}$.
For (\ref{eq:10}) they are even smaller due to the absorption
effects. 
For this reason the bremsstrahlung rate is sensitive only to 
the behavior of $\sigma(\rho)$ at $\rho\lsim 1/m_{e}\ll r_{B}$, 
where $r_{B}$ is the Bohr radius. 
We write the dipole cross section in the form
\beq
\sigma(\rho)=\rho^{2}C(\rho)\,,
\label{eq:15}
\eeq
where 
$
C(\rho)=Z^{2}C_{el}(\rho)+Z C_{in}(\rho)\,.
$
Here the terms $\propto Z^{2}$ and $\propto Z$ correspond to 
elastic and inelastic intermediate states in interaction
of $e^{+}e^{-}$ pair with an atom. 
For the atomic potential
$\phi(r)=4\pi(Z\alpha/r)\exp(-a/r)$
 $C_{el}(\rho)$ is given by \cite{BGZA2}
\bea
C_{el}(\rho)=8\pi\left(\frac{\alpha a}{\rho}\right)^{2}
\left[1-\frac{\rho}{a}K_{1}\left(\frac{\rho}{a}\right)\right]
\approx
4\pi\alpha^{2}
\left[\log\left(\frac{2a}{\rho}\right)+\frac{(1-2\gamma)}{2}\right]\,,
\;\;\gamma=0.577\,.
\label{eq:17}
\eea
For nuclei of finite radius $R_{A}$ 
$C_{el}(\rho\lsim R_{A})=C_{el}(R_{A})$.
At $\rho\ll r_{B}$ the factor $C_{in}(\rho)$ also can be
parametrized in form (\ref{eq:17}). We use the parameters
$a=0.83\, r_{B}Z^{-1/3}$ for the elastic component, and 
$a=5.2\, r_{B}Z^{-2/3}$ for the inelastic one.
This choice allows to reproduce the elastic and inelastic 
contributions to the Bethe-Heitler cross section evaluated 
in the standard approach with realistic atomic form factors \cite{Tsai}.
 
%***************************************************************
{\bf 4.}
In Fig. 1 we compare the results of calculations 
(solid curve) of  
the bremsstrahlung rate with the one measured in \cite{SLAC}
for a gold target with $L=0.7\%X_{0}\approx 0.023$ mm 
($X_{0}$ is the radiation length) and 25 GeV electron beam.
We also show the prediction of frozen-size 
approximation (\ref{eq:14}) (dashed curve),
the radiation rate obtained for the infinite medium
(long-dashed curve), and the Bethe-Heitler spectrum 
(dot-dashed curve). We have found that the normalization of 
the experimental spectrum disagrees with our theoretical 
prediction. The theoretical curves in Fig.~1 were
multiplied by the factor 0.93. This renormalization
brings the calculated spectrum in a good agreement with the data
of ref. \cite{SLAC}. The origin of the above disagreement 
in normalization is not clear. The authors of ref. 
\cite{SLAC} give the systematic error 3.2\%. However,
normalizations of the spectrum for
gold targets with $L= 0.7\%X_{0}$ and $L=0.1\%X_{0}$ 
at the photon momentum $k\sim 500$ MeV, where the 
LPM suppression is expected to be small, differ by $\sim 20-30$\%.

For 25 GeV electrons 
$l_{f}\approx 0.47\cdot (1 \mbox{MeV}/k(\mbox{MeV}))$ mm in the region of
$k$ shown in Fig.~1. One can conclude from this figure that the radiation
rate calculated using Eqs. (\ref{eq:4}),~(\ref{eq:9}),~(\ref{eq:10}) 
is close to the prediction of the frozen-size 
approximation (\ref{eq:14}) for the photons with $l_{f}\gsim L$, while for
the photons with $l_{f}\lsim L$ it is close to the spectrum 
for the infinite medium. 
To illustrate the role of the finite target thickness better we present 
in Fig.~2 the LPM suppression factor defined as $S=dP/dx/dP_{BH}/dx$
as a function of the ratio $h=L/l_{f}$ for several values of 
the photon momentum. The calculations were performed for a gold target and
25 GeV electron beam.
Fig.~2 demonstrates that the edge effects come into play
at $L\lsim l_{f}$. One can also see from Fig.~2 that for 
low photon momenta the edge effects vanish faster.
This fact is a consequence of a stronger suppression of the coherence
length in radiation of soft photons. 

Notice that the suppression factor has
a minimum at $L\approx l_{f}$ for 100 and 400 MeV photons. 
This minimum reflects the two-edge interference for a plate target.
One can expect a more pronounced interference effects
for structured targets. 
In Fig.~3 we show our results for the LPM suppression factor 
for a two segment gold target. 
Qualitatively our results for the interference effects are similar to those 
of ref. \cite{B}, in which the bremsstrahlung rate for structured targets
%interference effects were evaluated
was evaluated
modelling the  medium by the potential 
$
U(\r,z)=-\r\cdot \mbox{\bf E}_{\perp}(z)\,,
$
where $\mbox{{\bf E}}_{\perp}$ is a random transverse electric field
\cite{BD}.
However, for our realistic electron-atom
interaction the maxima and minima in the
spectra are less pronounced than for the model medium used
in ref.\cite{B}. For a 
homogeneous target our spectrum differs from obtained 
by Blankenbecler by $\sim 10$\%.

The reason for this disagreement is as follows.
Using the technique of refs. \cite{BGZA2,BGZLPM} one can
show that the model potential of refs. \cite{BD,B} translates 
in our approach to the following choice of the dipole
cross section
$$
\sigma(\rho)=\frac{2\pi \alpha\rho^{2}}{n}\int\limits_{-\infty}^{\infty}
d z \,\langle \mbox{\bf E}_{\perp}(0)\cdot
\mbox{\bf E}_{\perp}(z)
\rangle\,
$$
in which the important logarithmic $\rho$-dependence which 
derives from the Coulomb interaction is missed.
We conclude that the model of refs. \cite{BD,B} is too crude 
for a quantitative simulation of the LPM effect in a real medium.

To summarize, we evaluated the LPM effect
in QED for finite-size homogeneous and structured targets.
For the first time we performed a rigorous theoretical analysis of
the experimental data on the LPM effect obtained at SLAC \cite{SLAC}.
The theoretical predictions up to a normalization factor
0.93 are in a good agreement with the spectrum
measured at SLAC  \cite{SLAC} for the homogeneous gold target
with $L=0.7\%X_{0}$ and 25 GeV electron beam. For structured targets we predict
minima and maxima in the photon spectra.

I would like to thank B.Z.~Kopeliovich and N.N.~Nikolaev for discussions
and reading the manuscript.   
I am grateful to J.~Speth for the hospitality at KFA,
J\"ulich, where a part of this work was done.

%\end{references}

{\hspace*{-1.35cm}\Large{Figures:}}

\begin{enumerate}
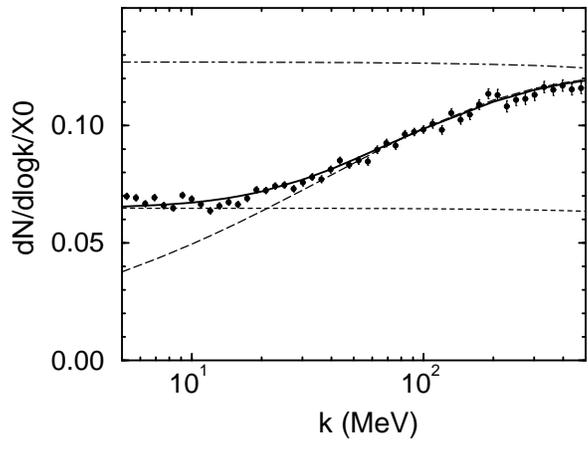
\item[Figure 1:]{
The bremsstrahlung spectrum for 25 GeV electrons incident
on a gold target with a thickness of $0.7\%X_{0}$.
The experimental data are from ref. \cite{SLAC}.
The full curve shows our results obtained using 
Eqs.~(\ref{eq:9}),~(\ref{eq:10}). The dashed curve was
obtained in the frozen-size approximation (\ref{eq:14}).
The long-dashed curve shows the spectrum for the infinite 
medium. The Bethe-Heitler spectrum is shown by the dot-dashed curve.
}
 
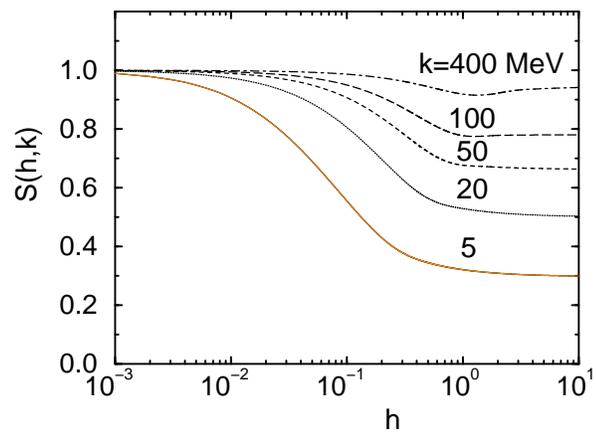
\item[Figure 2:]{
The LPM suppression factor for 25 GeV electron incident
on a homogeneous gold target as a function of the ratio
$h=L/l_{f}$ and the photon momentum.
}

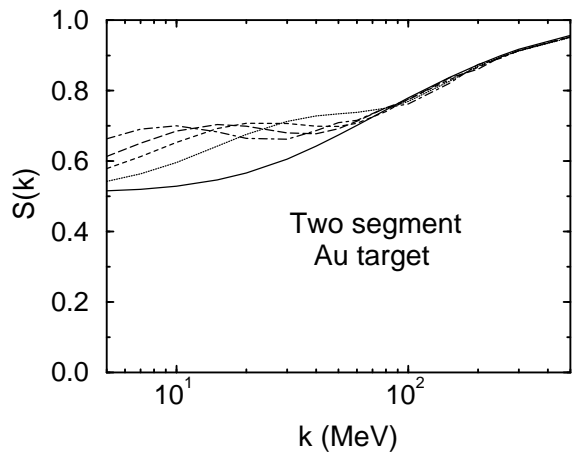
\item[Figure 3:]{
The LPM suppression factor for 25 GeV electron incident on
a two segment gold target. The thickness of each segment is $0.35\%X_{0}$.
The set of gaps is as follows: 0 (solid curve), 
$0.7\%X_{0}$ (dotted curve), $1.4\%X_{0}$ (dashed curve),
$2.1\%X_{0}$ (long-dashed curve), $3.5\%X_{0}$ (dot-dashed curve). 
}

\end{enumerate}
 
\end{document}